\documentclass{article}

\usepackage{graphicx}
\usepackage{hyperref}

\author{Hugo Kuijf}
\title{MixMicrobleedNet: segmentation of cerebral microbleeds using nnU-Net}
\date{5 August 2021}

\begin{document}

\maketitle

\section{Introduction}

Cerebral microbleeds are hypointense, small, and round or ovoid lesions \cite{Pantoni2010,Greenberg2009}; visible on magnetic resonance imaging (MRI) with gradient echo, T2*, or susceptibility weighted (SWI) imaging\cite{Wardlaw2013,DeGuio2016,Smith2019}. Assessment of cerebral microbleeds is mostly performed by visual inspection, using validated rating scales such as the Microbleed Anatomical Rating Scale (MARS)\cite{Gregoire2009} or Brain Observer MicroBleed Scale (BOMBS)\cite{Cordonnier2009}. In the past decade, prior to the rise of deep learning technology in medical image analysis\cite{Litjens2017}, semi-automated tools to assist with cerebral microbleed detection have been developed. These include techniques based on unified segmentation\cite{Seghier2011}, support-vector machines\cite{Barnes2011}, or the radial symmetry transform\cite{Kuijf2012,Kuijf2013b,W2013}. In the more recent years, owing to the great advances provided by deep learning techniques, the number of methods for fully automatic microbleed detection has increased considerably\cite{Dou2016a,S2019,T2021,Almasni2020,Zhang2017b}.

In this work, we explore the use of nnU-Net\cite{Isensee2020} as a fully automated tool for microbleed segmentation. This self-configuring deep learning-based semantic segmentation method has shown good performance in a number of international biomedical segmentation competitions\cite{Maier-Hein2018}, but has not been applied to the task of cerebral microbleed detection and segmentation.

\section{Material and methods}

\subsection{Data}
Data was provided by the ``Where is VALDO?'' challenge of MICCAI 2021 (\url{https://valdo.grand-challenge.org/}). It consisted of T1, T2, and T2* images of 72 subjects; all aligned in the T2*-space. A manual segmentation of microbleeds was provided for every subject as a binary image.

\subsection{Pre-processing}
The world coordinates of the T2* image was copied into the T1 and T2 images, because they did not exactly match for all subjects. Images were renamed to match the requirements of nnU-Net. No further pre-processing was performed.

\subsection{nnU-Net}
Different settings and configurations of nnU-Net were explored for this task. The final method consists of nnU-Net in the ``3D full resolution U-Net'' configuration trained on all data (fold = `all'). No post-processing options of nnU-Net were used.

\subsection{Post-processing}
A number of post-processing options was explored, to reduce the number of false positive detections by the nnU-Net. However, visual inspection of the results showed that most false positive detections are most likely true microbleeds, which were missed during the initial visual rating (see Results). This is very common in a difficult task like visual rating of microbleeds and happens often\cite{Kuijf2013h}. Therefore, no post-processing was applied to avoid removing possible true detections.

\section{Results}
\begin{figure}[tbh]
\centering
\includegraphics[width=\textwidth]{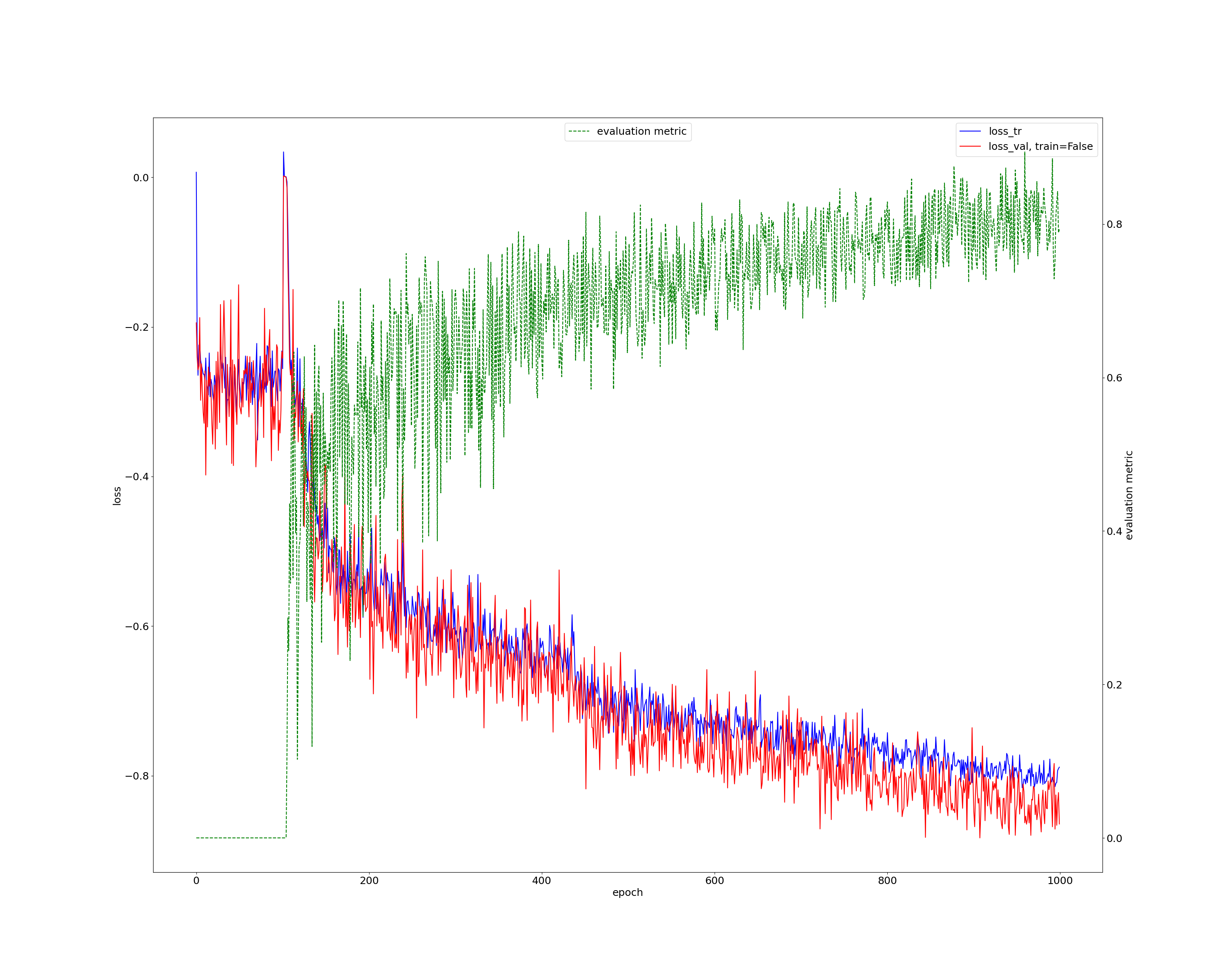}
\caption{Progress of nnU-Net during training.}
\label{fig:progress}
\end{figure}

Figure \ref{fig:progress} shows the progress of nnU-Net during training. The blue line is the training loss and the red line the validation loss. The green line shows the estimated Dice.

After training, nnU-Net reports a mean estimated Dice of 0.80 over all training cases. The reported False Discovery Rate (FDR) is 0.16 and the False Negative Rate (FNR) is 0.15.

\begin{figure}[tbh]
\centering
\includegraphics[width=0.7\textwidth]{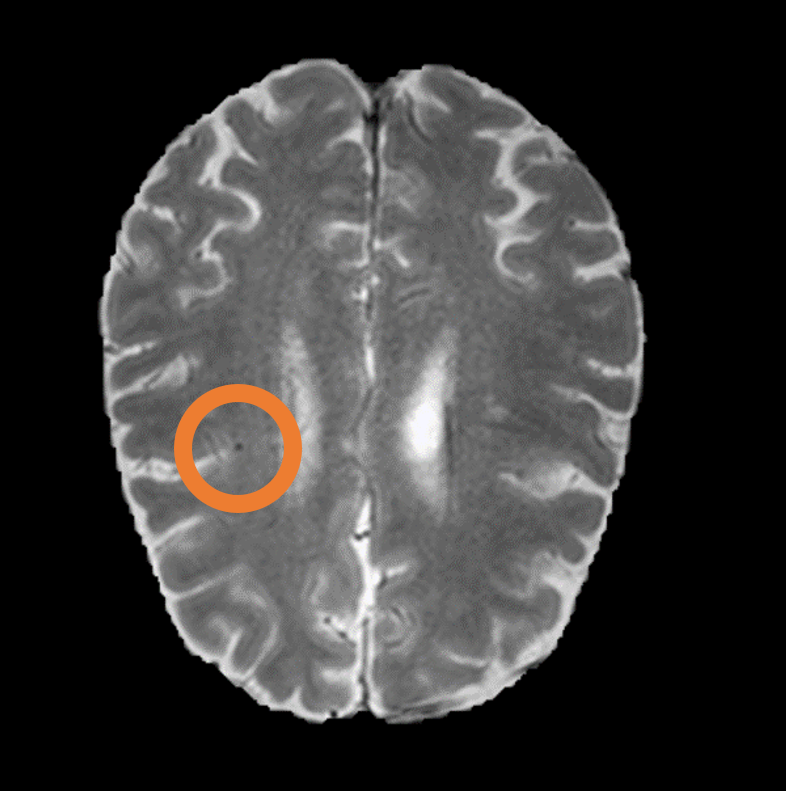}
\caption{Slice 23 of sub-101 showing a false positive detection that might actually be a true microbleed.}
\label{fig:slice23}
\end{figure}

Visual inspection of the results showed that most of the reported false positives could be an actual microbleed that might have been missed during visual rating. For example, this location (see Figure \ref{fig:slice23}) in sub-101 on slice 23.

\section{Discussion}

The organizers of the VALDO challenge posted the microbleed-task as a semantic segmentation task. However, owing to the blooming effect\cite{McAuley2011} of iron deposits on MR images, the visible size and volume of a microbleed is dependent on a number of factors; including field strength and echo time. Having a microbleed detection task, as opposed to a segmentation task, might therefore have been more logical choice. 

Taking that into account, we first explored detection methods that then would be followed by a segmentation method, to obtain results suitable for the challenge. A number of methods were explored, including the nnDetection\cite{Baumgartner2021} framework that recently won the Aneurysm Detection And segMentation (ADAM) challenge\cite{Timmins2021}. Intermediate empirical results showed that a detection method followed by a segmentation method might not outperform nnU-Net for this task.

\section{More information}
Source code is available at: \url{https://github.com/hjkuijf/MixMicrobleedNet}. The docker container hjkuijf/mixmicrobleednet can be pulled from \url{https://hub.docker.com/r/hjkuijf/mixmicrobleednet}.

\bibliographystyle{ieeetr} 
\bibliography{}

\end{document}